\documentstyle[12pt,aps,prd,preprint]{revtex}
\begin{document}
\title{Evidence for a null entropy of extremal black holes}
\author{Shahar Hod}
\address{The Racah Institute for Physics, The
Hebrew University, Jerusalem 91904, Israel}
\date{\today}
\maketitle

\begin{abstract}
We present some arguments in support of a {\it zero} entropy for {\it extremal}
black holes. These rely on a combination of both quantum,
thermodynamic, and statistical physics arguments. 
This result may shed some light on the nature of these extreme
objects. In addition, we show that within a {\it quantum} framework 
the capture of a particle by an
initially extremal black hole always results with a final nonextremal black hole. 
\end{abstract}
\bigskip

\section{introduction}\label{introduction}
Extremal black holes have an important and controversial status in
black-hole physics. It had been traditionally believed that an extremal black hole is the
limiting case of its nonextremal counterpart; when the inner Cauchy
horizon and the outer event horizon coincide, the nonextremal black
hole becomes an extremal one \cite{MTW}. However, this traditional
point of view has been recently challenged by Hawking {\it et al.}
\cite{HaHoRo}, who based their arguments on the qualitative
differences between the topologies of extremal and nonextremal black
holes, differences which raise doubts about limiting arguments.

While it is well established that a nonextremal
black hole bears an entropy which is proportional to its surface 
area $S_{BH}=A/4\hbar$ \cite{Beken1,Haw1} (we use gravitational
units in which $G=c=1$), there is no general
agreement on the entropy of extremal black holes. 
Based on the different topologies of extremal and nonextremal
Reissner-Nordstr\"om black holes, Hawking {\it et al.}
\cite{HaHoRo} and Teitelboim \cite{Teit} argued that extremal black
holes have {\it zero} entropy
even though their event horizon has nonzero area. For further reading
see e.g., \cite{ZasGhoMit,Mit,WaSu,KieLou} and references therein. 

In this paper we examine the consistency of the Bekenstein-Hawking
area-entropy relation $S_{BH}=A/4\hbar$ with the properties of extremal black holes. To this end,
we shall use a combination of both quantum, thermodynamic, and
statistical physics arguments: In Sec. \ref{Sec2} we construct a 
gedanken experiment in which the (quantum) generalized second
law of thermodynamics is shown to be incompatible with the
area-entropy relation as applied to extremal black holes.  In Sec. \ref{Sec3} we show that the standard
{\it quantization} of angular momentum and electric charge in nature, when
applied to extremal black holes, are incompatible with the
area-entropy relation. The same argument leads to a zero-entropy
conjecture for extremal black holes. 

\section{Gedanken experiments with extremal black holes}\label{Sec2}

We consider a neutral object 
which is lowered towards an extremal Kerr-Newman black hole. 
We challenge the validity of accepted physical laws in 
the most `dangerous' situation, i.e., when the energy delivered to the
black hole is as small as possible. We therefore bring the object as
close to the horizon as possible, and then drop it in. 
The descent of the body, if sufficiently slow, is known to be an
adiabatic process which causes no change in the black-hole horizon area
\cite{TPM,Beken2,Mayo}.

To {\it zeroth} order in particle-hole interaction 
the energy (energy-at-infinity) ${\cal E}^{(0)}$ of the object in the black-hole 
spacetime is given by Carter's \cite{Carter}
integrals (constants of motions). As first shown by Christodoulou
\cite{Chris} (see also \cite{ChrisRuff}), 
${\cal E}^{(0)}(r=r_+)=\Omega^{(0)}L_z$ at the point of capture, where 
$\Omega^{(0)}=a/(r^2_++a^2)$ is the angular velocity of the black
hole, $L_z$ is the conserved angular momentum of the particle, 
and $r_+=M$ is the location of the black-hole horizon.

One should also consider {\it first}-order interactions between the black
hole and the object's angular momentum: As the particle spirals into the
black hole (in the case $L^2_z \neq 0$) 
it interacts with the black hole, so the horizon generators 
start to rotate, such that at the point of assimilation the black-hole
angular velocity $\Omega$ has changed from $\Omega^{(0)}$ to
$\Omega^{(0)}+\Omega^{(1)}_c$. The corresponding first-order energy
correction is ${\cal E}^{(1)}=\Omega^{(1)}_cL_z$. 
On dimensional analysis we expect $\Omega^{(1)}_c$ to be
of the order of $O(L_z/M^3)$. In fact, Will \cite{Will} has performed a
perturbation analysis for the problem of a ring of particles rotating
around a slowly rotating (neutral) black hole, and found
$\Omega^{(1)}_c=L_z/4M^3$. As would be expected from a perturbative
approach, $\Omega^{(1)}_c$ is proportional to $L_z$. To our best
knowledge, no exact calculation of $\Omega^{(1)}_c$ has been performed 
for generic Kerr-Newman black holes. We therefore write 
${\cal E}^{(1)}=\omega L^2_z$, and obtain

\begin{equation}\label{Eq1}
{\cal E}={\cal E}^{(0)}+{\cal E}^{(1)}={{aL_z} \over {M^2+a^2}}+ \omega L^2_z\  ,
\end{equation}
for the particle's energy at the point of capture.

The assimilation of the object results with a change $\Delta
M={\cal E}$ in the black-hole mass, and a change $\Delta J=L_z$ in its
angular momentum. With the plausible assumption of cosmic 
censorship \cite{Pen} one may 
argue from Hawking's area theorem \cite{Haw2} ($\Delta A \geq 0$) to find

\begin{equation}\label{Eq2}
\omega \geq {M \over {2(M^2+a^2)^2}}\  .
\end{equation}
For the analysis to be self-consistent, the black-hole condition
$M^2-a^2-Q^2 \geq 0$ should be satisfied after the assimilation of the
object. This requires 

\begin{equation}\label{Eq3}
\omega \geq {{M(M^2-3a^2)} \over {2(M^2+a^2)^3}}\  ,
\end{equation}
which is always a weaker condition than the condition Eq. (\ref{Eq2}). We therefore
conclude, that provided cosmic censorship is respected, the final
black hole (in the case $L^2_z \neq 0$) is {\it not} extremal [the two expressions Eq. (\ref{Eq2})
and Eq. (\ref{Eq3}) coincide for the unique case $a=0$ (first
considered in \cite{Need}), in which case
{\it second}-order interactions between the black
hole and the object's angular momentum should be considered].

The increase in black-hole surface area due to the assimilation of the
object is of the order of $O(|L_z|)$. Evidently, this can be minimized
for $L_z=0$, in which case one finds $\Delta A=0$. This result is
consistent with Hawking's ({\it classical}) area theorem \cite{Haw2}.

We next examine the gedanken experiment from the point of view of a {\it quantum}
theory of gravity. A complete quantum theory of gravity is, of 
course, beyond our present reach. We do have, however, an important fingerprint
of the elusive theory. The Bekenstein-Hawking area-entropy relation
$S_{BH}=A/4(\hbar G/c^3)^{1/2}$ involves the universal constants
$\hbar$ and $G$ of quantum theory and gravitation, respectively.
It therefore allows a glance into the realm of quantum
gravity. The concept of black-hole entropy is intimately related to the 
generalized second law (GSL) of thermodynamics \cite{Beken1,Beken3} 
``The sum of the black-hole entropy and the common (ordinary) entropy
in the black-hole exterior never decreases''. 

We realize that the result $\Delta S_{BH}=0$ (this is a direct
consequence of the result $\Delta A=0$ provided the relation
$S_{BH}=A/4\hbar$ holds true for extremal black holes) 
does not respect the GSL; the object's entropy disappears with no obvious 
physical mechanism to compensate for its loss. We recall, however,
that the essence of a {\it quantum} theory is 
the Heisenberg quantum uncertainty principle. It implies that
$\delta L^2_z$ cannot vanish identically for a fairly localized object, since
according to the uncertainty principle this would give rise to a large
uncertainty in its canonically conjugate variable, the azimuthal angle
$\phi$. [Recall for example, that the stationary states of the hydrogen atom, which
are also eigenstates of the $L_z$ operator (and hence have a definite
value of $L_z$, and a vanishing $\delta L^2_z$), have a probability
distribution which is $\phi$-{\it independent}, i.e., the
corresponding wave function is completely unlocalized in the $\phi$
direction, whereas the semiclassical analysis 
requires the descending object to be fairly localized.] Specifically, we have the
uncertainty relation

\begin{equation}\label{Eq4}
\sqrt {\delta L^2_z} \geq {\hbar \over {2 \delta \phi}} \gtrsim \hbar\  ,
\end{equation}
where $\delta \phi \ll 1$ if the particle is in the equatorial plane,
and $\delta \phi$ can be made of the order of unity if the particle is
near the black-hole poles (note that the
particle cannot be localized at the pole itself, since this would
violate the uncertainty relation between the angle variable $\theta$,
and the corresponding canonically conjugate angular momentum).

Taking cognizance of Eq. (\ref{Eq4}) we 
obtain $(\Delta A)_{min}=O(\hbar)$ within the quantum framework. 
According to the standard area-entropy relation, the corresponding
increase in black-hole entropy is of the order of
$O(1)$. However, this increase in black-hole entropy cannot
guarantee the GSL's validity; the disappeared entropy 
(the object's entropy) is generically larger than $O(1)$. Thus, assuming the
validity of the area-entropy relation for extremal black holes one
finds $\Delta S_{tot}= \Delta S_{BH} -S_{object}< 0$ in our gedanken 
experiment, in contradiction with the GSL \cite{note1}. This motivates 
the conjecture that the area-entropy relation is not applicable for extremal black holes.

The important point to be emphasized is, that within the {\it quantum}
theory the final black hole (obtained after the capture of the object) is
{\it not} extremal any more [see the discussion after
Eqs. (\ref{Eq2}) and (\ref{Eq3})]. The final black-hole
entropy is therefore given by the standard Bekenstein-Hawking relation 
$S^{fin}_{BH}=A^{fin}/4\hbar$. Assuming that the initial black-hole
entropy is zero (for an extremal black hole), one obtains
$\Delta S_{tot}= \Delta S_{BH} -S_{object}=A^{fin}/4\hbar -S_{object}
>0$ (for black holes much larger than the test object, as required by
the semiclassical analysis), in agreement with the GSL.

The argument presented in this section supports the idea that the
area-entropy relation is not valid for extremal black holes. On the
other hand, the zero-entropy conjecture for extremal black holes 
was shown to be compatible 
with the GSL. [Of course, this is not to say that a zero entropy is the only 
resolution; an entropy of the form (say) 
$S=\ln (A/\hbar)$ for extremal black holes would also be
compatible with the GSL. However, the gedanken experiment reveals that
the standard proportionality between black-hole surface area and entropy,
if applied to extremal black holes would contradict the GSL.] 
In the next section we give further evidence in support of 
the conjecture that extremal black holes have zero entropy.

\section{Angular-momentum and charge quantization vs. the area-entropy
  relation}\label{Sec3}

The quantization of extremal black holes was discussed by Mazur
\cite{Mazur} and Bekenstein (see, e.g., \cite{Beken5}): The extremal
Kerr black hole is defined by the relation $M^4=J^2$, which implies
$A=8\pi M^2$. One enforces the quantization by replacing the total angular-momentum
by the well-known eigenvalues of the corresponding quantum operator,
namely $J^{2} \to j(j+1) \hbar^{2}$, where $j$ is a non-negative 
integer or half-integer. 
One therefore obtains 

\begin{equation}\label{Eq5}
A_{j}=8 \pi \sqrt{j(j+1)} \hbar \  ,
\end{equation}
for the area eigenvalues of the extremal Kerr black hole. 

In the spirit of Boltzmann-Einstein formula in {\it statistical physics}
one  relates $exp(S_{BH})$ to the number of microstates of the
black hole that correspond to a particular external macrostate. Thus, 
the {\it thermodynamic} relation $S_{BH}=A/(4 \hbar)$ 
between black-hole surface area and entropy implies that the
degeneracy corresponding to the $j$th area level is 

\begin{equation}\label{Eq6}
g_{j}=exp[2 \pi \sqrt{j(j+1)}]\  .
\end{equation}
This quantity is, however, {\it not} an integer. 
We therefore conclude that the area-entropy {\it thermodynamic}
relation, if applied to extremal black holes, is not compatible with 
a combination of {\it quantum} and standard {\it statistical physics} 
arguments (namely, the Boltzmann-Einstein formula). 
[This state of affairs should be contrasted with the corresponding situation 
for non extremal black holes \cite{Beken5,Hod3}, where area quantization, 
statistical physics arguments, and the Bekenstein-Hawking
thermodynamic relation all agree !]

The corresponding discussion for the extreme Reissner-Nordstr\"om black hole is very
similar to the one presented for extremal Kerr black holes. 
The extreme Reissner-Nordstr\"om 
black hole is defined by the relation 
$|Q|=M$, which implies $A=4 \pi Q^2$ for the black-hole surface
area. The quantization of its area eigenvalues 
was discussed by Mazur \cite{Mazur} and Bekenstein (see, e.g.,
\cite{Beken5}); one enforces the quantization by 
replacing $Q \to qe$, where $q$ is an integer and $e$ is the
elementary charge. One thus obtains 

\begin{equation}\label{Eq7}
A_{q}=4 \pi \alpha q^{2} \hbar \  ,
\end{equation}
for the area eigenvalues of the extremal Reissner-Nordstr\"om black
hole, where $\alpha =e^2/\hbar$ is the fine-structure constant. 
The area spectrum Eq. (\ref{Eq7}), togather with the thermodynamic 
relation $S_{BH}=A/(4 \hbar)$, imply that the degeneracy corresponding to the $q$th
area eigenstate is 

\begin{equation}\label{Eq8}
g_{q}=exp(\pi \alpha q^{2})\  ,
\end{equation}
which is, again, not an integer. We therefore recover our previous 
conclusion, that for extremal black holes {\it quantum} and {\it statistical} physics arguments 
are not compatible with the area-entropy {\it thermodynamic} relation.

Similar analysis reveals the fact, that for an extreme Kerr-Newman black
hole (a synthesis of the former two extreme black holes) with or
without a magnetic monopole, the area-entropy thermodynamic
relation is inconsistent with quantum and statistical physics
arguments. This supports the idea that extremal black holes do not comply with 
the standard area-entropy relation. 
Moreover, taking cognizance of Eqs. (\ref{Eq5}) and (\ref{Eq7}) one
finds that the entropy of extremal black holes should equal a constant
(the logarithm of an integer) in order to be 
compatible with the standard statistical physics interpretation of
entropy. 

We should comment, however, that the entropy of an extremal black hole
could agree with the Bekenstein-Hawking entropy, provided it is not
interpreted as a statistical (Boltzmann) entropy.

\section{Summary}\label{Sec4}

We have shown that the Bekenstein-Hawking thermodynamic relation, if applied
to extremal black holes, leads to violation of the generalized
second law of thermodynamics. This motivates the conjecture that the
standard area-entropy relation is not valid for extremal black holes,
as first suggested (from a completely different point of view) by
Hawking {\it et al.} \cite{HaHoRo}, and by Teitelboim
\cite{Teit}. 

Moreover, we have shown that in a quantum framework, 
the assimilation of a particle by an extremal black hole always
results with a final {\it non} extremal
black hole (under the plausible assumption of cosmic censorship). This
final result, togather with a null entropy for extremal black holes
restore the validity of the GSL to our gedanken experiment 
because the huge increase in black-hole entropy (from zero to $A/4\hbar$) 
compensate for the loss of the object's entropy.

We have further shown that for extremal black holes the area-entropy
{\it thermodynamic} relation is inconsistent 
with {\it quantum} and {\it statistical} physics arguments. The later
imply that the entropy of extremal black holes should equal a
constant. This should be contrasted with the corresponding situation
in the physics of non extremal black holes,
where area quantization, statistical physics arguments, and the
Bekenstein-Hawking area-entropy thermodynamic relation are all 
compatible \cite{Beken5,Hod3}.

We finally note that a zero entropy for extremal black holes is in
agreement with the interpretation of black-hole entropy as the
logarithm of the number of quantum mechanically distinct ways in which
the particular black hole could have been made through successive 
excitations \cite{ZuTh,Muk}, because it is has been shown that a non extremal black
hole cannot be transformed into an extremal one \cite{Wang,Hod4} (and
the present paper also reveals, that within a quantum framework 
an extremal black hole cannot be transformed into another extremal
black hole). These results therefore imply 
$S_{ext}=\ln 1=0$ for the entropy of extremal black holes.

\bigskip
\noindent
{\bf ACKNOWLEDGMENTS}
\bigskip

I thank Jacob D. Bekenstein and Avraham E. Mayo for discussions.
This research was supported by a grant from the Israel Science Foundation.


\begin{thebibliography}{99}

\bibitem{MTW} C. W. Misner, K. S. Thorne, and J. A. Wheeler, {\it Gravitation} (Freeman, San Francisco, 1973).

\bibitem{HaHoRo} S. W. Hawking, G. Horowitz, and S. Ross, Phys. Rev. D {\bf 51}, 4302 (1995).

\bibitem{Beken1} J. D. Bekenstein, Ph.D. thesis, Princeton 
University,1972 (unpublished); Lett. Nuov. Cim. {\bf 4}, 737 (1972); 
Phys. Rev. D {\bf 7}, 2333 (1973).

\bibitem{Haw1} S. W. Hawking, Commun. Math. Phys. {\bf 43}, 199 (1975).

\bibitem{Teit} C. Teitelboim, Phys. Rev. D {\bf 51}, 4315 (1995); 
{\bf 52}, 6201(E) (1995).

\bibitem{ZasGhoMit} O. B. Zaslavskii, Phys. Rev. Lett. {\bf 76},
  2211 (1996); A. Ghosh and P. Mitra, Phys. Rev. Lett. {\bf 78},
  1858 (1997).

\bibitem{Mit} P. Mitra, Phys. Lett. B {\bf 441}, 89 (1998).

\bibitem{WaSu} B. Wang and R. K. Su, Phys. Rev. D {\bf 59}, 104006 (1999).

\bibitem{KieLou} C. Kiefer and J. Louko, Annalen Phys. {\bf 8}, 67 (1999).

\bibitem{TPM} K. S. Thorne, R. H. Price and D. A. 
MacDonald, {\it Black Holes: The Membrane Paradigm} (Yale University
Press, London, 1986).

\bibitem{Beken2} J. D. Bekenstein, in {\it Black Holes, Gravitational
    Radiation and the Universe}, eds. B. R. Iyer and B. Bhawal (Kluwer,
    Dordrecht 1998).

\bibitem{Mayo} A. E. Mayo, Phys. Rev. D {\bf 58}, 10400 (1998).

\bibitem{Carter} B. Carter, Phys. Rev. {\bf 174}, 1559 (1968).

\bibitem{Chris} D. Christodoulou, Phys. Rev. Lett. {\bf 25}, 1596
  (1970).

\bibitem{ChrisRuff} D. Christodoulou and R. Ruffini, Phys. Rev. {\bf
    D4}, 3552 (1971).

\bibitem{Will} C. M. Will, Astrophys. J. {\bf 191}, 521 (1974).

\bibitem{Pen} R. Penrose, Riv. Nuovo Cimento {\bf 1}, 252 (1969);
  R. Penrose in 
{\it General Relativity, an Einstein Centenary Survey},
eds. S. W. Hawking and W. Israel (Cambridge University Press, 1979). 

\bibitem{Haw2} S. W. Hawking, Phys. Rev. Lett. {\bf 26}, 1344 (1971).

\bibitem{Need} T. Needham, Phys. Rev. D {\bf 22}, 791 (1980).

\bibitem{Beken3} J. D. Bekenstein, Phys. Rev. D {\bf 9}, 3292 (1974).

\bibitem{note1} This state of affairs should
be {\it contrasted} with the corresponding one involving nonextremal
black holes. For non extremal black holes, the increase in black-hole surface
area (entropy) ensures the validity of the GSL \cite{Beken4,Hod1,BekenMay,Hod2}.

\bibitem{Beken4} J. D. Bekenstein, Phys. Rev. D {\bf 23}, 287 (1981).

\bibitem{Hod1} S. Hod, Phys. Rev. D {\bf 61}, 024018 (2000); e-print
  gr-qc/9901035.

\bibitem{BekenMay} J. D. Bekenstein and A. E. Mayo, Phys. 
Rev. D {\bf 61}, 024022 (2000); e-print gr-qc/9903002.

\bibitem{Hod2} S. Hod, Phys. Rev. D {\bf 61}, 024023 (2000); e-print
  gr-qc/9903010, gr-qc/9903011.

\bibitem{Mazur} P. Mazur, Gen. Rel. Grav. {\bf 19}, 1173 (1987).

\bibitem{Beken5} J. D. Bekenstein in XVII Brazilian National Meeting
  on Particles and Fields, eds. A. J. da Silva et. al. (Brazilian
  Physical Society, Sao Paulo, 1996); Proceedings of the VIII 
Marcel Grossmann Meeting on General Relativity, eds. T. Piran and 
R. Ruffini (World Scientific , Singapore, 1998).

\bibitem{Hod3} S. Hod, Phys. Rev. Lett. {\bf 81}, 4293 (1998).

\bibitem{ZuTh} W. H. Zurek and K. S. Thorne, Phys. Rev. 
Lett. {\bf 54}, 2171 (1985).

\bibitem{Muk}V. Mukhanov, JETP letters {\bf 44}, 63 (1986).

\bibitem{Wang} B. Wang, R. K. Su, P. K. N. Yu and E. C. M. Young, 
Phys. Rev. D {\bf 57}, 5284 (1998).

\bibitem{Hod4} S. Hod, Phys. Rev. D {\bf 60}, 104031 (1999).

\end{thebibliography}
\end{document}